\documentclass[prd,preprint,floatfix,nofootinbib,tightenlines,preprintnumbers,superscriptaddress]{revtex4-2}

\usepackage[english]{babel}
\usepackage{amsmath,amssymb,bm,slashed}
\usepackage{graphicx}
\usepackage[sort&compress]{natbib}
\usepackage[usenames,dvipsnames,svgnames]{xcolor}
\usepackage[normalem]{ulem}
\usepackage{hyperref}
\usepackage[capitalize]{cleveref}
\crefformat{footnote}{#2\footnotemark[#1]#3}
\crefname{paragraph}{Paragraph}{Paragraphs}
\definecolor{red}{rgb}{1.0, 0, 0}
\definecolor{green}{rgb}{0.0,0.7,0.2}

\usepackage[section]{placeins}

\usepackage{pifont}

\hypersetup{
  colorlinks,
  linktoc=page,
  linkcolor=blue,
  citecolor=ForestGreen
}

\allowdisplaybreaks

\definecolor{baic}{rgb}{0.00392156862745098, 0.45098039215686275, 0.6980392156862745}
\definecolor{bpic}{rgb}{0.8705882352941177, 0.5607843137254902, 0.0196078431372549}
\definecolor{paic}{rgb}{0.00784313725490196, 0.6196078431372549, 0.45098039215686275}
\definecolor{ppic}{rgb}{0.8352941176470589, 0.3686274509803922, 0.0}
\definecolor{indv}{rgb}{0.8, 0.47058823529411764, 0.7372549019607844}
\definecolor{naive}{rgb}{0.5803921568627451, 0.5803921568627451, 0.5803921568627451}
\definecolor{indv2}{rgb}{0.71, 0.71, 0.71}
\definecolor{indv_2state}{rgb}{0.984313725490196, 0.6862745098039216, 0.8941176470588236}

\makeatletter\AtBeginDocument{\let\@elt\relax}\makeatother
\usepackage{extarrows}
\usepackage{float}
\floatstyle{plaintop}
\usepackage[caption=false]{subfig} 
\usepackage{array}
\usepackage{multirow}

\hypersetup{
  colorlinks,
  linktoc=page,
  linkcolor=blue,
  citecolor=ForestGreen
}

\allowdisplaybreaks

\bibpunct{[}{]}{,}{n}{}{,}

\newcommand{\diag}{\text{diag}}

\renewcommand{\vec}[1]{{\mathbf{#1}}}

\providecommand{\para}[1]{\left(#1\right)}
\providecommand{\bracket}[1]{\left[#1\right]}
\providecommand{\abs}[1]{\left\lvert#1\right\rvert}

\newcommand{\beq}{\begin{equation}}
\newcommand{\eeq}{\end{equation}}

\newcommand{\pr}{p}

\newcommand{\yd}{\{y\}}
\newcommand{\Ndof}{\ensuremath{N_{\textrm{dof}}}}

\DeclareMathOperator{\perf}{perf}
\DeclareMathOperator{\subspace}{sub}
\DeclareMathOperator{\C}{C}
\DeclareMathOperator{\K}{K}

\DeclareMathOperator{\T}{T}

\DeclareMathOperator{\AIC}{AIC}

\DeclareMathOperator{\BAIC}{BAIC}

\DeclareMathOperator{\IC}{IC}

\usepackage{amsthm}

\begin{document}
\title{Model averaging approaches to data subset selection}

\author{Ethan T.~Neil}              \email[Email: ]{ethan.neil@colorado.edu}
\affiliation{Department of Physics, University of Colorado, Boulder, CO 80309, USA}

\author{Jacob W.~Sitison}          \email[Email: ]{jacob.sitison@colorado.edu}
\affiliation{Department of Physics, University of Colorado, Boulder, CO 80309, USA}

\date{\today} 

\begin{abstract}

Model averaging is a useful and robust method for dealing with model uncertainty in statistical analysis.  Often, it is useful to consider data subset selection at the same time, in which model selection criteria are used to compare models across different subsets of the data.  Two different criteria have been proposed in the literature for how the data subsets should be weighted.  We compare the two criteria closely in a unified treatment based on the Kullback-Leibler divergence, and conclude that one of them is subtly flawed and will tend to yield larger uncertainties due to loss of information.  Analytical and numerical examples are provided.

\end{abstract}


\maketitle

\newpage
\section{Introduction}
\label{sec:intro}

In data analysis, one is often tasked with assessing a set of candidate models' ability to match data and then use these models to make predictions (e.g., of future data or physical parameters).  A simple approach is model selection, where a single model is selected from the available candidates based on some data-driven criteria.  A related alternative to model selection is model averaging, where each model is assigned a probability weight with which a composite model is constructed via a weighted average of the individual models.  The model probability weights are often determined by adapting model selection criteria; hence, we will refer to such model comparison problems generically as \emph{model variation}.

A commonly used model variation approach is based on the Akaike information criterion (AIC):
\begin{align}
\AIC=-2\ln L^*+2k,
\end{align}
where $L^*$ is the maximum value of the likelihood function and $k$ is the number of parameters in the model considered.  The AIC will be discussed in detail in \cref{sec:review}, but for now it suffices to know that models that minimize the AIC are favored in model variation.

A related problem to model variation is that of data subset variation, where one seeks a subset of the data to be fitted. 
There has been some recent disagreement in the literature on how to adapt model variation criteria to data subset variation.  More specifically, two different forms of the AIC for this use case have been proposed:
\begin{align}
\AIC^{\subspace}&=-2\ln L^*+2k-d_{\K},\\
\AIC^{\perf}&=-2\ln L^*+2k-2d_{\K},
\end{align}
where $d_{\K}$ is the number of fitted data points.  We refer to the methods by which these formulas are derived as the ``subspace'' method and the ``perfect model'' method, respectively, as described in \cref{sec:data-subset-variation}. The ways in which these formulas are derived are quite different.  The subspace formula was derived in \cite{Borsanyi:2020mff} (with some results from \cite{Borsanyi:2014jba}) from a frequentist perspective.  On the other hand, the perfect model formula was derived in \cite{Jay:2020jkz} within a Bayesian framework.  Direct comparison of the two derivations is made more difficult by their very different contexts, even though the two approaches agree on the form of the AIC when $d_{\K}$ is held fixed (up to subtle differences in the definition of the log-likelihood in Bayesian versus frequentist formulations).

A more direct way to compare the two derivations was made apparent in \cite{Neil:2022joj}, which reformulates the Bayesian model averaging framework proposed in \cite{Jay:2020jkz} using the Kullback-Leibler (K-L) divergence (see \cref{subsec:kl-div} below) as the fundamental object of interest.  Since the derivation of the subspace formula in \cite{Borsanyi:2020mff} is also based on the K-L divergence, this allows us to compare both formulas directly.  Our conclusion will be that $\AIC^{\subspace}$ does not give rise to meaningful comparisons, whereas $\AIC^{\perf}$ does.  Formally, $\AIC^{\subspace}$ compares distributions on different measurable spaces whereas $\AIC^{\perf}$ keeps the underlying measurable space fixed.  

While the two methods perform similarly in many cases, $\AIC^{\subspace}$ will tend to favor fits with less data (i.e., larger data cuts) compared to $\AIC^{\perf}$, which can result in inflated statistical uncertainties.  It may be tempting to conclude from this behavior that the $\AIC^{\subspace}$ formula is simply a more conservative choice.  In some sense this is true, but not in the form of a bias-variance tradeoff \cite{Mehta2019}.  Rather, we argue that the $\AIC^{\subspace}$ uncertainties will in some cases be inflated unnecessarily but with no compensating reduction in bias.

The paper is outlined as follows.  First, in \cref{sec:review} we briefly review key ingredients in model variation.  In \cref{subsec:kl-div}, we introduce the Kullback-Leibler (K-L) divergence as a means of comparing two distributions.  We connect the K-L divergence to the problem of data subset variation in \cref{sec:data-subset-variation}.  In \cref{sec:ex}, we show an example of comparing distributions with the K-L divergence, and then study numerical examples of data subset averaging to further compare the subspace and perfect-model constructions.  \cref{sec:conclusion} summarizes our results.  \cref{sec:app-asymptotic-kl} shows some analytic results useful for derivation of the $\AIC^{\subspace}$, while \cref{sec:AIC-dof} explores the consequences of rewriting both AIC formulas in terms of the number of degrees of freedom.

\section{Review of model variation formalism}
\label{sec:review}
In this section, we briefly review some of the key ideas in model variation and establish our notation.  We largely follow the presentation of \cite{Neil:2022joj}, to which the reader is referred for further details.

Our discussion of model variation is always taken in the context of fitting models to data.  We define a model space $\{ M_\mu \}$, which may or may not contain the ``true model'' $M_{\T}$ from which the data arise.\footnote{For simplicity of the current discussion, we assume only a single true model; more complicated scenarios are discussed in \cite{Neil:2022joj}.}  The true model defines a probability distribution $\pr_{M_{\T}}(z) \equiv \pr(z | M_{\T}) = \pr(z)$ from which the future data $z$ (in general, a vector of length $d$) are drawn.  Model variation then relies on determination of how ``close'' a given model $M_\mu$ is to the true model $M_{\T}$.

The distribution over all future data $z$ is only available formally; in practice, inference is done based on a finite data set $\yd$ with sample size $N$ drawn from $\pr(z)$.  The fundamental object of interest for model variation is then the sample likelihood $\pr(\yd | M_\mu)$.  By convention, it is common to work with this quantity in the form of an ``information criterion'' (IC), which takes the generic form
\begin{align}
\IC_{\mu}\equiv-2\ln\pr(\yd|M_{\mu}).
\end{align}
There are a variety of ways in which estimators for $\IC_{\mu}$ can be constructed, particularly with regard to parametric models that depend on some parameter vector $\vec{a}$.  For simplicity, we will focus on the commonly-used Akaike information criterion (AIC), which uses a ``plug-in'' estimate with the best-fit parameter point $\vec{a}^*$ obtained from the data sample $\yd$:
\begin{align}
\AIC_{\mu}\equiv-2\ln\pr(\yd|M_{\mu},\vec{a}^*)+2k,
\end{align}
where $k$ is the number of parameters in the vector $\vec{a}$.  The second term is an asymptotic bias correction, which in some sense may be thought of as correcting for ``double use'' of the data in first obtaining $\vec{a}^*$ and then $\pr(\yd|M_{\mu},\vec{a}^*)$; see \cite{Jay:2020jkz,Neil:2022joj} for further details and for discussion of the assumptions required for use of the $\AIC$.  For other ways to deal with parameter dependence, resulting in different information criteria, see \cite{Neil:2022joj}.

In applications within lattice field theory, it is most common to use least-squares fitting for analyzing models in this way.  In this case, the sample likelihood is given by the standard chi-squared statistic,
\begin{align}
-2\ln\pr(\yd|M_{\mu},\vec{a}^*)=\sum_{i=1}^N\chi_i^2(\vec{a}^*)=(N-1)d+\hat{\chi}^2(\vec{a}^*),
\end{align}
where $N$ is the number of observations and $d$ is the dimension of the data space.  (The $(N-1)d$ term is typically constant over the data space, and hence may be ignored.  This term will not always be constant in \cref{sec:data-subset-variation}, so we will write it explicitly when relevant.  For a derivation of this term, see Appendix B of \cite{Jay:2020jkz}.)  The expression above is written in terms of both the individual and average chi-squared statistics, 
\begin{align}
\chi_i^2(\vec{a})=(y_i-f_{M_{\mu}}(\vec{a}))^T\Sigma^{-1}(y_i-f_{M_{\mu}}(\vec{a})),\\
\hat{\chi}^2(\vec{a})=(\bar{y}-f_{M_{\mu}}(\vec{a}))^T\hat{\Sigma}^{-1}(\bar{y}-f_{M_{\mu}}(\vec{a})),
\end{align}
where $\Sigma$ is the covariance matrix, $\hat{\Sigma}=\Sigma/N$ is the standard error covariance matrix, $f_{M_{\mu}}(a)$ is the model function, and $\bar{y}=\frac{1}{N}\sum_iy_i$ is the sample mean.  

Thus in the case of least-squares, we have that
\begin{align}
\AIC_{\mu}&=(N-1)d+\hat{\chi}^2(\vec{a}^*)+2k.
\end{align}
This formula may be used for comparing different models, assuming that the data set $\yd$ is held fixed.  Variation of the data set as well, which is the main focus of this paper, will be discussed in \cref{sec:data-subset-variation} below.  In order to set the stage for that discussion, we next turn to the Kullback-Leibler divergence as a more fundamental way of comparing $\pr(z | M_\mu)$ with the true distribution $\pr_{M_{\T}}(z)$.

\subsection{Kullback-Leibler divergence}
\label{subsec:kl-div}

The Kullback-Leibler (K-L) divergence was introduced in \cite{Kullback1951} as means of assessing the ``distance'' between two distributions.  Let $f(z)$ and $g(z)$ be two probability density distributions on the same measurable space.  The K-L divergence between these distributions is defined by\footnote{While $I(f,g)$ is referred to as the K-L divergence or K-L information interchangeably in the model variation literature, \cite{Kullback1951} originally refers to $I(f,g)$ as ``information'' whereas the ``divergence'' is defined as $$J(f,g)=I(f,g)+I(g,f)=\int dF(z)\ln\frac{f(z)}{g(z)}-\int dG(z)\ln\frac{g(z)}{f(z)}=\int dz(f(z)-g(z))\ln\frac{f(z)}{g(z)}.$$  We adhere to the modern usage of ``divergence'' rather than the historical nomenclature.}
\begin{align}
I(f,g)\equiv\int dF(z)\ln\frac{f(z)}{g(z)}=\int dzf(z)\ln\frac{f(z)}{g(z)}.
\end{align}
The K-L divergence satisfies $I(f,g) \geq 0$, vanishing if and only if $f$ and $g$ are equivalent in the sense of distributions (i.e., $f(z)=g(z)$ almost everywhere).  An example of how to make comparisons with the K-L divergence is provided in \cref{subsec:kl-ex}.

In the discussion to follow, we will draw on two useful results from \cite{Kullback1951}, namely Theorems 3.1 and 4.1.  While we will not state these formally here as they are rather technical, we will try to give some intuition for their content.

Theorem 3.1 states that the K-L divergence is additive for independent random events.  For example, if we have separable joint probability densities $f(x,y)=f_x(x)f_y(y)$ and $g(x,y)=g_x(x)g_y(y)$, then we can write the K-L divergence as
\begin{align}
I(f,g)=I(f_x,g_x)+I(f_y,g_y),
\end{align}
which is straightforward to verify from the definition of $I(f,g)$.

Theorem 4.1 explains how the K-L divergence changes under ``onto'' (i.e., surjective) transformations of the underlying measurable space.  Specifically, if $\mathcal{O}$ is an onto operator, $f_{\mathcal{O}}(z)=f(\mathcal{O}z)$, and $g_{\mathcal{O}}(z)=g(\mathcal{O}z)$, then
\begin{align}
I(f_{\mathcal{O}},g_{\mathcal{O}})\leq I(f,g),
\end{align}
with equality if and only if $\mathcal{O}$ is sufficient.\footnote{Informally, a sufficient statistic is one that contains the maximum amount of information available about the underlying distribution \cite{larsen2005introduction}.}  In words, the K-L divergence $I(f,g)$ in some sense measures the amount of information $g$ contains about $f$; the theorem then states that the K-L divergence is reduced if any information is lost under $\mathcal{O}$, or that it remains the same if $\mathcal{O}$ retains all the original information (i.e. if it is sufficient).

\section{Data subset variation}
\label{sec:data-subset-variation}
We now consider the extension of the model variation formalism of \cref{sec:review} to the problem of data subset variation.  The goal is to select $d_{\C}$ dimensions of the $d$-dimensional data $y_i$ to be ignored, and fit models only to the other $d_{\K}\equiv d-d_{\C}$ dimensions. (The subscript ``$\C$" refers to the ``cut" portion of the data and the subscript ``$\K$" refers to the ``kept" portion of the data.)  Here we describe and compare the two constructions that lead to two different criteria, $\AIC^{\subspace}$ and $\AIC^{\perf}$ (introduced in \cref{sec:intro}.)

Before doing so, we introduce some additional notation following the connection of K-L divergence and information criteria developed in \cite{Neil:2022joj}.  We will specialize to the K-L divergence comparing the true distribution to a given model, which is the central quantity of interest of model variation:
\begin{align} \label{eqn:model-variation-KL}
I_{\T}(M_{\mu}) \equiv I(\pr_{M_{\T}},\pr_{M_{\mu},\vec{a}^*})=E_z[\ln\pr_{M_{\T}}(z)]-E_z[\ln\pr(z|M_{\mu},\vec{a}^*)],
\end{align}
where $E_z[\dots]$ denotes the expectation with respect to the true distribution $\pr_{M_{\T}}(z)$.  Here we have specialized to the case of the ``plug-in'' K-L divergence that leads to the AIC as described in \cref{sec:review}.  We have suppressed the $\vec{a}^*$-dependence of $I_{\T}(M_{\mu})$ for notational simplicity.

For a fixed data set, the first term of the K-L divergence is constant over the model space, and is thus ignored when defining information criteria such as the AIC. However, now that we are considering data variation, we must revisit the first term as it will no longer be constant in all cases.  As we will see in \cref{subsec:aic-subspace,subsec:aic-perf}, the construction of the $\AIC^{\subspace}$ will require treating a non-constant first term, whereas the construction of the $\AIC^{\perf}$ retains a constant first term.

The second term is estimated by the information criterion divided by a conventional normalizing factor of $-2N$,
\begin{equation}\label{eqn:aic-definition}
E_z[\ln\pr(z|M_{\mu},\vec{a}^*)] \simeq -\frac{1}{2N} \AIC_\mu,
\end{equation}
where ``$\simeq$" denotes an asymptotically unbiased estimate.

While somewhat oversimplified,\footnote{For example, two different data subsets of the same dimension could be considered, as in the case of allowing both ``minimum'' and ``maximum'' cuts on the data to vary simultaneously.  Since the penalty term of interest only depends on the dimension of the subset, our discussion generally applies to this more complicated case as well.} it is sufficient for our purposes to label different candidate data subsets with the dimension of the kept data space $d_{\K}$ and hence write the K-L divergence as $I_{\T}(M_{\mu},d_{\K})$.  While this notation will have different meanings for the two methods considered, the meaning of this shorthand will be clear from context.

In both constructions, in the limit of vanishing correlations between the cut and kept data spaces, by Theorem 3.1 of \cite{Kullback1951} we may divide the total K-L divergence into two terms: 
\beq \label{eqn:aic-division}
I_{\T}(M_\mu, d_{\K}) = I_{\T,\C} + I_{\T,\K}(M_\mu),
\eeq
where $I_{\T,\C}$ and $I_{\T,\K}$ are defined on the cut and kept data spaces, respectively.  Note that by definition, the cut space does not depend on the model of interest $M_\mu$ at all.  

The assumption of vanishing correlations may seem severe in some applications.  In the perfect model construction, although this assumption is useful for yielding the simplest forms of information criteria, it can be relaxed in principle; see \cite{Neil:2022joj} for details.  On the other hand, for the subspace construction, this assumption appears to be essential; we see no obvious path to dealing with cut-kept data correlations in this approach.

\subsection{Subspace construction}
\label{subsec:aic-subspace}

We first consider the ``subspace'' form of the AIC for data subset variation, which was proposed in \cite{Borsanyi:2020mff}, using key results from  \cite{Borsanyi:2014jba}.  The basic idea of $\AIC^{\subspace}$ is simple: for a given data cut, we omit the cut portion of the data completely,
\beq
I_{\T}^{\subspace}(M_\mu, d_{\K}) \equiv I_{\T,\K}(M_\mu)
\eeq
or in other words $I_{\T,\C}^{\subspace} = 0$.  There is still explicit dependence on $d_{\K}$, arising from keeping both terms in the K-L divergence \cref{eqn:model-variation-KL}.  In our notation, and switching to the AIC, we have
\begin{align}
\label{eqn:aic-subspace} \AIC_{\mu,d_{\K}}^{\subspace}&=(N-1)d_{\K}+\hat{\chi}_{\K}^2(\vec{a}^*)+2k-Nd_{\K}=\hat{\chi}_{\K}^2(\vec{a}^*)+2k-d_{\K},
\end{align}
where $\hat{\chi}_{\K}^2(\vec{a}^*)$ is evaluated only over the kept portion of the data.  Compared to the standard definition of the AIC as an estimator of the second term in the K-L divergence, as in \cref{eqn:aic-definition}, the subspace AIC includes an additional $-Nd_{\K}$ term arising from the first term in the K-L divergence, which is no longer constant under the subspace construction; see \cref{sec:app-asymptotic-kl} for a brief derivation and a discussion of the required assumptions.

While the derivation of  $\AIC^{\subspace}$ is mathematically sound, we argue that it does not provide a legitimate means of comparing different data subsets via the K-L divergence.   By allowing the dimension of $z$ to change, the subspace construction no longer compares models to $\pr_{M_{\T}}(z)$ over the same space.  This violates the spirit of model selection, the goal of which is to assess the ability of each model to predict future data $z$ generated from $\pr_{M_{\T}}(z)$ \cite{Konishi2008}.  In the subspace construction, we are comparing model predictions for different subsets of $z$ to one another directly, which is not meaningful.  This can lead to counter-intuitive results, a simple example of which is shown in \cref{subsec:kl-ex}.

More concretely, we can demonstrate that $\AIC^{\subspace}$ has an undesirable tendency to cut data too aggressively, all else being equal.  Consider a fixed model $M_{\mu}$ on a set of data $\yd$ with each observation partitioned into $y_i=(y_{i,\C},y_{i,\K})$.  For simplicity, we assume $y_{i,\C}$ are independent from $y_{i,\K}$.  The K-L divergences as estimated by the $\AIC^{\subspace}$ for the full data and for the cut data are given by
\begin{align}
I_{\T}^{\subspace}(M_{\mu},d)&\simeq-\frac{1}{2N}\para{\hat{\chi}_{\C}^2(\vec{a}^*)+\hat{\chi}_{\K}^2(\vec{a}^*)+2k-d_{\C}-d_{\K}},\\
I_{\T}^{\subspace}(M_{\mu},d_{\K})&\simeq-\frac{1}{2N}\para{\hat{\chi}_{\K}^2(\vec{a}^*)+2k-d_{\K}},
\end{align}
respectively. Define $\mathcal{O}$ such that
\begin{align}
y_{i,\K}=\mathcal{O}y_i,
\end{align}
i.e., $\mathcal{O}=\begin{pmatrix}0_{\C}&1_{\K}\end{pmatrix}$ where $0_{\C}$ and $1_{\K}$ are the $d_{\C}\times d_{\C}$ and $d_{\K}\times d_{\K}$ zero and identity matrices, respectively.  Since $\mathcal{O}$ is onto, we see from Theorem 4.1 of \cite{Kullback1951} (see discussion in \cref{subsec:kl-div}) that
\begin{align}
\label{eqn:thm-4.1}
I_{\T}^{\subspace}(M_{\mu},d)\geq I_{\T}^{\subspace}(M_{\mu},d_{\K}).
\end{align}
So, asymptotically, the $\AIC^{\subspace}$ will never choose to fit all of the data over fitting some of the data. (We emphasize that this result is asymptotic and will not always be the case at finite $N$.  In fact, we will see that this condition is explicitly violated in our numerical tests below, although the qualitative behavior of over-cutting data is still evident.)

We have shown that the $\AIC^{\subspace}$ will asymptotically fail to favor fits with more data, under the right conditions (it will still do well at rejecting models that describe the data poorly, in which case the $\hat{\chi}^2$ term will dominate).  By simply discarding part of the K-L divergence, the subspace construction inevitably results in loss of information.  This information loss is an obstruction to meaningful comparisons in model variation; in essence, the discarded information reduces the K-L divergence in the same way that a better model would, and so rewards data cuts somewhat too heavily.  The effect is subtle, since there is still some penalty for cutting data in $\AIC^{\subspace}$.  However, we will see in further examples below that the lost information leads to larger uncertainties on statistical estimates obtained with $\AIC^{\subspace}$.

\subsection{Perfect model construction}
\label{subsec:aic-perf}

To avoid the problems that arise from comparing K-L divergences across different spaces, as in \cref{subsec:aic-subspace}, an alternative approach is to define a K-L divergence that discriminates between different data cuts in a way that holds the full data set fixed. Here again, the basic idea is simple: rather than discarding the cut-data K-L divergence in \cref{eqn:aic-division}, we keep both terms but add a model which describes the cut data,
\begin{align}
I_{\T}^{\rm piece-wise}(M_{\mu},d_{\K}) \equiv I_{\T,\C}(M_{\mu}^{\C})+I_{\T,\K}(M_{\mu}),
\end{align}
In this construction, the true model $M_{\T}$ on the entire data space is compared to a piece-wise model $M_\mu^{\rm piece-wise} = (M_{\mu}^{\C},M_{\mu})$.  The effect of choosing the data cut $d_{\K}$ is to change the relative size of the two component models.

Completing this approach requires us to define a model $M_\mu^{\C}$ for the cut data.  This may seem counterproductive at first glance, since the purpose of cutting data away is to ignore it and carry out inference focused on the kept data and $M_\mu$.  However, while the choice of $M_\mu$ will depend on context, we are free to select a general $M_\mu^{\C}$ that minimizes the impact of $I_{\T,\C}$ as much as possible.

One such construction proposed in \cite{Jay:2020jkz} and studied further in \cite{Neil:2022joj} is to choose $M_{\mu}^{\C}$ such that the $\hat{\chi}_{\C}^2(\vec{a}^*)$ vanishes, referred to in the references as a ``perfect'' model $M_{\mu}^{\perf}$. As an example, $M_{\mu}^{\perf}$ could be a polynomial interpolant of $\bar{y}_{\C}$.  Overfitting the cut data space in this manner comes with the cost of adding $d_{\C}$ additional parameters to the composite model $M_{\mu}^{\rm piece-wise}$.  This leads to a bias correction of $2d_{\C}$, following from the $2k$ fit parameter penalty in the AIC.\footnote{Because the perfect model is very specialized, one might wonder whether it violates some of the assumptions used in deriving the standard AIC and thus should have a different bias correction.  We believe that this is not the case, and the correction of $+2d_{\C}$ is right; to verify this, we have done numerical tests in a simple Gaussian example showing this bias explicitly.  The numerical tests are included as a Jupyter notebook in our code repository, referenced in \cref{subsec:num-ex} below.} Therefore,
\begin{align}
I_{\T,\C}(M_{\mu}^{\perf})\simeq-\frac{1}{2N}[(N-1)d_{\C}+2d_{\C} - Nd_{\C}]=-\frac{1}{2N}d_{\C},
\end{align}
and thus 
\begin{align}
I_{\T}^{\perf}(M_{\mu},d_{\K})&\simeq-\frac{1}{2N}\bracket{\hat{\chi}_{\K}^2(\vec{a}^*)+2k+d_{\C}-d_{\K}}\\
&=-\frac{1}{2N}\bracket{\hat{\chi}_{\K}^2(\vec{a}^*)+2k-2d_{\K}+d}.
\end{align}
The resultant form of the AIC is thus
\begin{align} 
\AIC_{\mu,d_{\K}}^{\perf}&=\hat{\chi}_{\K}^2(\vec{a}^*)+2k-2d_{\K} + d. \label{eqn:aic-perfect}
\end{align}
The $d$ term is an overall constant and will therefore drop out when comparing models with the $\AIC^{\perf}$ (e.g., when evaluating the normalized model weights).

In the derivation of $\AIC^{\perf}$ above, we have included the $-Nd_{\C}$ and $-Nd_{\K}$ factors from the first term in the K-L divergence above (see \cref{sec:app-asymptotic-kl}), which required some assumptions to derive.  We have done so only for the sake of comparison to \cref{subsec:aic-subspace}; for the perfect model construction, the first term in the K-L divergence is always an overall constant that may be ignored in $\AIC^{\perf}$.  Note that we are also ignoring some terms that arise from the normalization of the model probability densities; these either cancel or lead to overall constants that drop out of model variation.  See the references or the discussion of similar normalizing terms in \cref{sec:app-asymptotic-kl}.

The end result is quite similar to $\AIC^{\subspace}$ in \cref{eqn:aic-subspace} except that the penalty for cutting data is stronger: $-2d_{\K}$ as opposed to $-d_{\K}$.  This additional factor accounts for the information contained in the $I_{\T,\C}$ term describing the cut data.  In particular, we emphasize that in this picture of data subset selection, Theorem 4.1 of \cite{Kullback1951} does not apply; because the data space is held fixed and the difference between data cuts amounts to changing the model itself, there is no transformation $\mathcal{O}$ that maps $I_{\T}^{\perf}(M_\mu, d)$ onto $I_{\T}^{\perf}(M_\mu, d_{\K})$.  Unlike the subspace construction, this allows the perfect model construction to favor keeping all of the data when all else is equal.

\subsection{Some analytic examples}
\label{subsec:analytic-comp}

In order to give further context for the similarities and differences between $\AIC^{\subspace}$ and $\AIC^{\perf}$, here we consider some brief examples where they can be studied analytically.

First, let $M_1$ be a model that does not approach the true model $M_{\T}$ and contains $k$ parameters.  Since $M_1\not\rightarrow M_{\T}$ in the limit of large sample size $N$, $\hat{\chi}_1^2=O(N)$.  Therefore,
\begin{align}
\AIC_{1,d_{\K}}^{\subspace}&=\hat{\chi}_{1}^2 + 2k -d_{\K}\approx\hat{\chi}_{1}^2,\\
\AIC_{1,d_{\K}}^{\perf}&=\hat{\chi}_{1}^2 + 2k -2d_{\K}\approx\hat{\chi}_{1}^2.
\end{align}
In both cases, the $\AIC$ is $O(N)$.  Hence, $M_1$ will be rejected when compared to a model that does approach $M_{\T}$.  This shows that the difference between the two data subset variation methods is subleading, and the goodness of fit is the dominant contribution in both formulas.

Next, let $M_2$ be the true model $M_{\T}$ with zero free parameters.\footnote{We choose zero parameters for simplicity here; a parametric model with $k$ parameters that converges to the true model will show exactly the same qualitative difference between subspace and perfect model constructions in the large $N$ limit.}  In this case, it can be shown that $\hat{\chi}_2^2\rightarrow d_{\K}$ in the large $N$ limit, and thus
\begin{align}
\AIC_{2,d_{\K}}^{\subspace}&=\hat{\chi}_{2}^2-d_{\K}\rightarrow0,\\
\AIC_{2,d_{\K}}^{\perf}&=\hat{\chi}_{2}^2-2d_{\K}\rightarrow-d_{\K}.
\end{align}
Here, the perfect model construction will choose to retain the maximum amount of information from the data and keep all data points.  Meanwhile, the subspace construction is agnostic to the amount of data fit.  This will put more weight on smaller data sets, which can lead to larger statistical uncertainties when doing inference.

We emphasize that the above examples, while hopefully useful for illustrating the differences between $\AIC^{\subspace}$ and $\AIC^{\perf}$, are somewhat idealized and do not apply generally.  In order to see how the two variations on the $\AIC$ perform in a more realistic example, we now turn to numerical study.

\section{Numerical examples}
\label{sec:ex}

\subsection{Distribution comparisons with the K-L divergence}
\label{subsec:kl-ex}

Before considering a realistic example of data subset variation, we begin with a simpler example of a K-L divergence--based comparison to demonstrate the subtleties of Theorem 4.1 (discussed in \cref{subsec:kl-div}.)  This example will demonstrate that comparison of K-L divergences computed against different spaces can lead to counter-intuitive results.

Consider the Gaussian distribution
\begin{align}
f(z)=\frac{1}{2\pi\sqrt{\abs{\Sigma}}}\exp\para{-\frac{1}{2}z^T\Sigma^{-1}z},
\end{align}
where $z = (z_1, z_2)$ is a two-dimensional vector and the covariance matrix $\Sigma=\begin{pmatrix}1&0.5\\0.5&2\end{pmatrix}$.  Suppose our goal is to estimate the variance of $f$ in the $z_1$ direction, which is (in the true model) $\sigma_1^2=1$.  To do so, we consider two candidate distributions:
\begin{align}
g(z)&=\frac{1}{2\pi\sqrt{\abs{\Sigma_g}}}\exp\para{-\frac{1}{2}z^T\Sigma_g^{-1}z},\\
h(z_1)&=\frac{1}{\sqrt{2\pi\sigma_h^2}}\exp\para{-\frac{1}{2}\frac{z_1^2}{\sigma_h^2}},
\end{align}
where $\Sigma_g=\diag\para{1.1,2}$ and $\sigma_h^2=1.5$.  Such candidates could, for example, be constructed by fitting to data drawn from $f$.  It is clear that $g$ better estimates $\sigma_1^2$ than $h$.

To see if the K-L divergence leads us to this conclusion, we must first determine how to compare $h$ to $f$.  Naively, one option would be to define
\begin{align}
f_1(z_1)&=\frac{1}{\sqrt{2\pi\sigma_1^2}}\exp\para{-\frac{1}{2}\frac{z_1^2}{\sigma_1^2}},
\end{align}
which is $f$ marginalized over the $z_2$ subspace. In the language of Theorem 4.1, we have defined the transformation $\mathcal{O}=\diag\para{1,0}$, which is onto but not sufficient. Therefore,
\begin{align}
I(f,g)&=\int dF(z)\ln\frac{f(z)}{g(z)}=0.069\dots,\\
I(f_1,h)&=\int dF_1(z)\ln\frac{f_1(z_1)}{h(z_1)}=0.036\dots,
\end{align}
which implies that $h$ is a better estimate than $g$.  So, the K-L divergence would lead us to use $\sigma_1^2\approx\sigma_h^2=1.5$ as our estimate for the $z_1$ variance instead of $\sigma_{g,1}^2=1.1$.

In the above comparison, the K-L divergence did not lead to the optimal estimate of $\sigma_1^2$.  This is a result of $g$ and $h$ being defined over different measurable spaces.  By projecting away the $z_2$ dependence of $I(f_1,h)$, the K-L divergence has been unduly reduced as explained by Theorem 4.1.  To properly compare $h$ to $f$, we need to extend the definition $h$ to the full measurable space on which $f$ is defined.  For example, we could define
\begin{align}
h^{\prime}(z)=\frac{1}{2\pi\sqrt{\abs{\Sigma_h}}}\exp\para{-\frac{1}{2}z^T\Sigma_h^{-1}z},
\end{align}
where $\Sigma_h=\diag\para{\sigma_h^2,\sigma_{g,2}^2}=\diag\para{1.5,2}$.  $h^{\prime}$ is the joint distribution of $h$ with $g_2$, $g_2$ being $g$ marginalized over $z_1$.  Now, we compare $I(f,g)$ to
\begin{align}
I(f,h^{\prime})&=\int dF(z)\ln\frac{f(z)}{h^{\prime}(z)}=0.103\dots,
\end{align}
and conclude that $g$ is a better estimate than $h^{\prime}$ giving as $\sigma_1^2\approx\sigma_{g,1}^2=1.1$ expected.

Alternatively, we also could have marginalized $g$ over $z_2$ and compared $g_1$ and $h$ to $f_1$.  This would have lead to the same conclusion that $\sigma_1^2\approx\sigma_{g,1}^2=1.1$.  In practice, it is typically better to instead extend the underlying measurable space (e.g., compare $g$ to $h^{\prime}$ rather than $g_1$ to $h$) as this retains more information from the data.  An exception is when a subspace of the data is excessively noisy, in which case throwing out some data from the analysis entirely can be advantageous (see \cite{Neil:2022joj} for further discussion).  With either approach, it is important that the underlying measure (e.g., $dF(z)$ or $dF_1(z)$) remain fixed across the set of candidates.

\subsection{Model averaging with subset selection}
\label{subsec:num-ex}
\begin{figure}[!htbp]
\centering
\includegraphics[width=\textwidth]{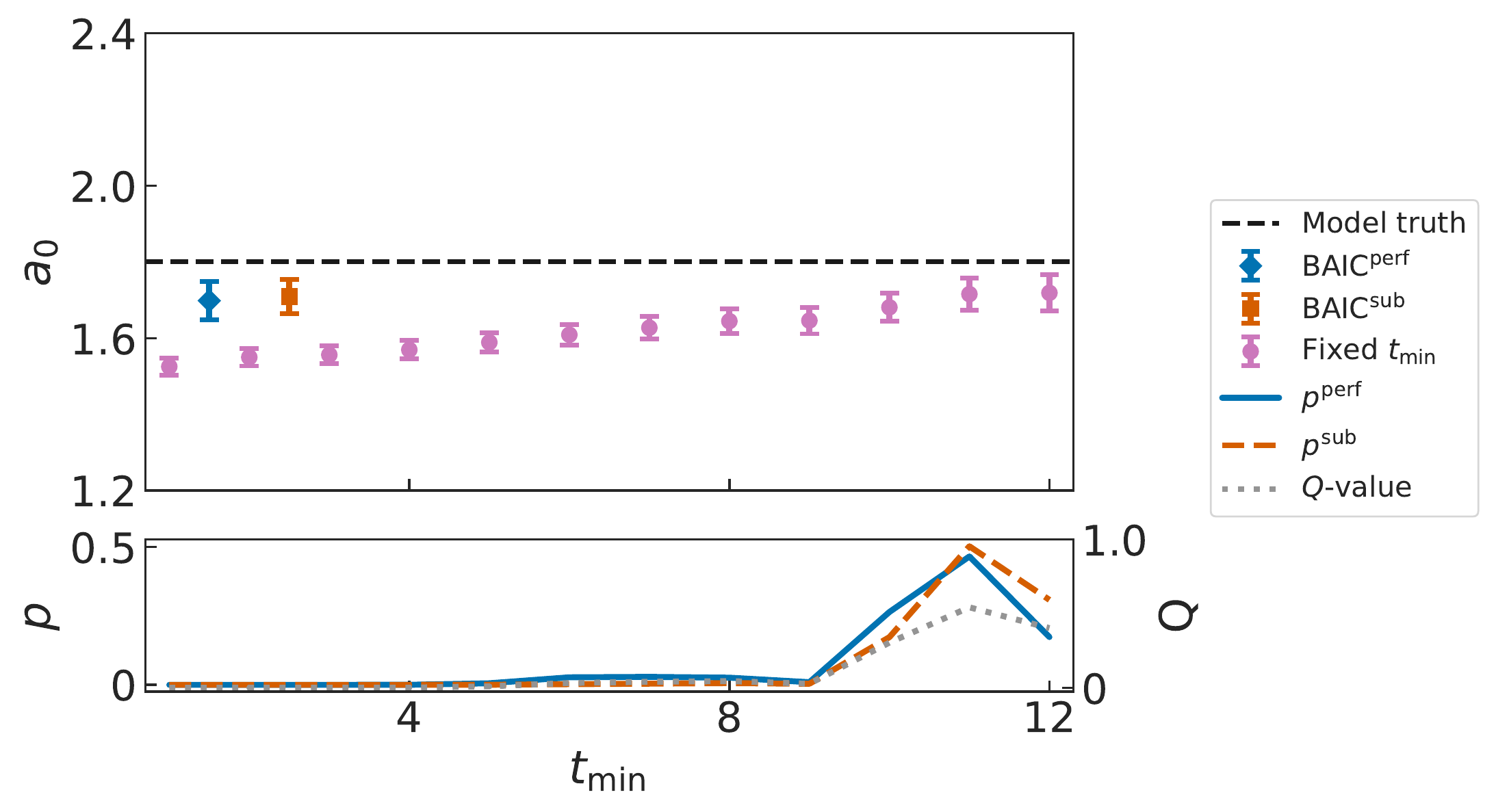}
\includegraphics[width=\textwidth]{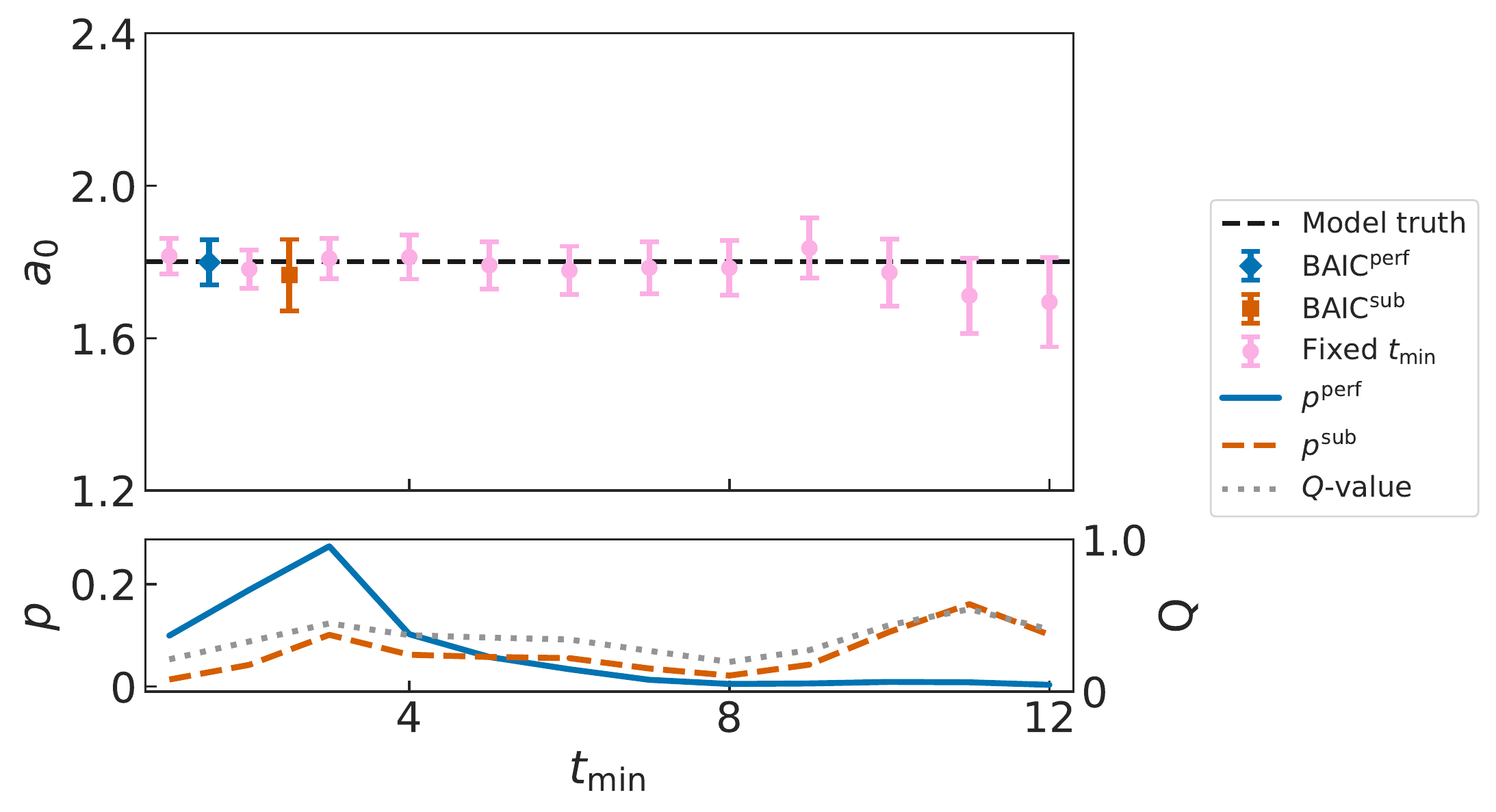}
\caption{Fit results for the intercept $a_0$ with fit models $f_0$ (top) and $f_1$ (bottom).  The top insets show the individual fit results from the data (with $N=320$) over data subsets $t\in[t_{\min},15]$ (purple {\color{indv}\LARGE\textbullet} for $f_0$ and pink {\color{indv_2state}\LARGE\textbullet} for $f_1$) compared to the true value $a_0=1.80$ (black dashed line) and model averaged results with $\BAIC^{\perf}$ (blue {\color{baic}\ding{117}}) and $\BAIC^{\subspace}$ (red {\color{ppic}$\blacksquare$}).  The bottom inset shows the standard $Q$-value (grey dotted line) and model weights $\pr(M_{\mu}|\yd)$ corresponding to the $\BAIC^{\perf}$ (blue solid curve) and $\BAIC^{\subspace}$ (red dashed curve).}
\label{fig:fixed_N}
\end{figure}

\begin{figure}[!htbp]
\centering
\includegraphics[width=0.8\textwidth]{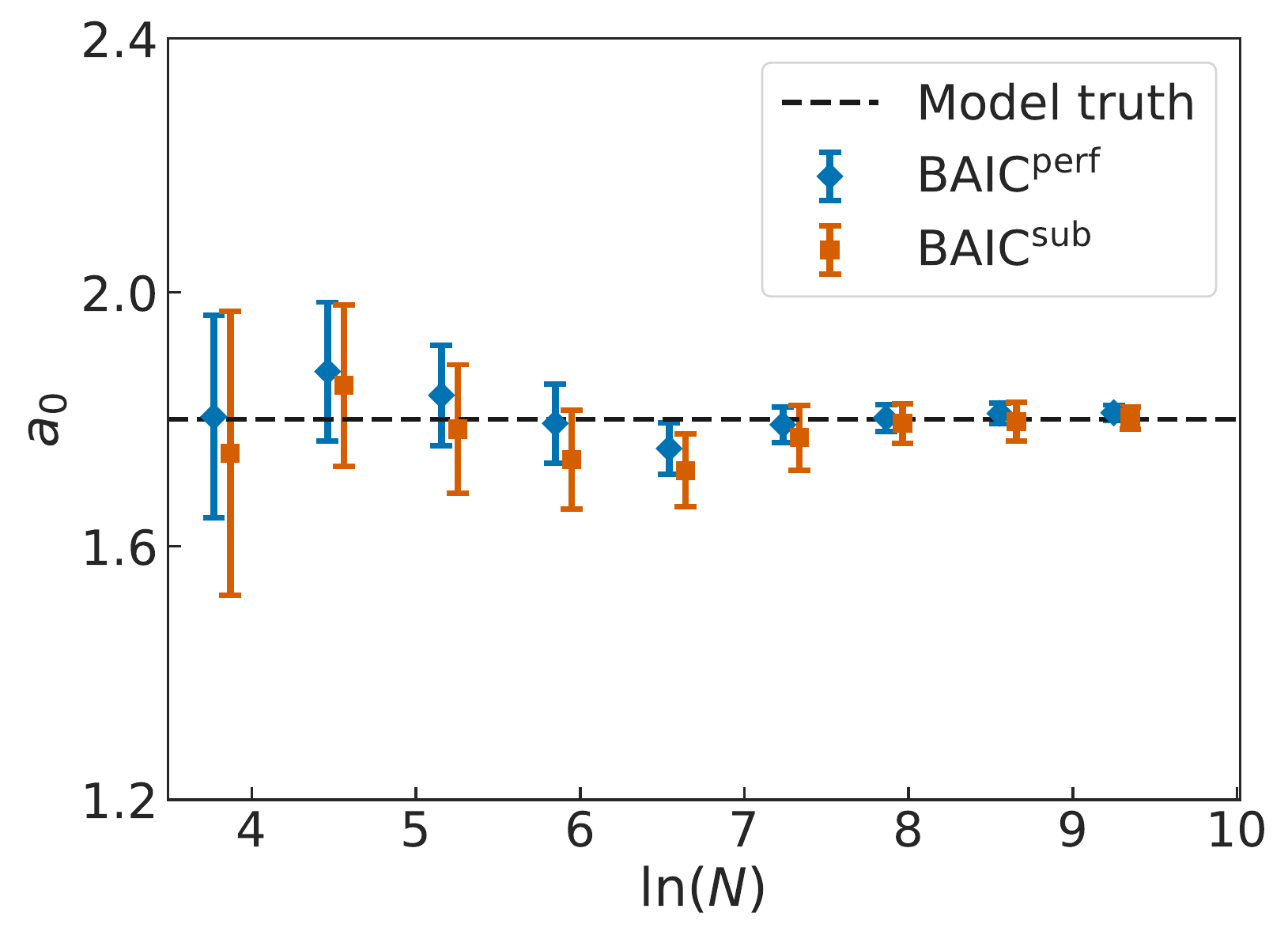}
\caption{$N$-dependent scaling of grand-averaged estimates (including constant and linear polynomials over all candidate data subsets) of the intercept $a_0$. Model averaged results with $\BAIC^{\perf}$ (blue {\color{baic}\ding{117}}) and $\BAIC^{\subspace}$ (red {\color{ppic}$\blacksquare$}) compared to the the true value $a_0=1.80$ (black dashed line) are shown.
}
\label{fig:N_scaling_ga}
\end{figure}

Here we show a numerical example of data subset averaging to compare the performance of $\BAIC^{\perf}$ to $\BAIC^{\subspace}$ when averaging over data subsets. The Bayesian AIC (BAIC) as defined in \cite{Neil:2022joj} is a Bayesian analog of the frequentist AIC.  The only difference is posterior mode is used for the best-fit parameters $\vec{a}^*$ in the BAIC as opposed to the maximum likelihood estimate in the AIC.  If the parameter priors are sufficiently diffuse, there are no significant theoretical or practical differences between the AIC and BAIC.  All Bayesian least squares fits are performed using the \texttt{lsqfit} package in Python \cite{Lepage2002,lsqfitGitHub}, which uses the Gaussian random variable data type from \texttt{gvar} \cite{gvarGitHub}.\footnote{The code used for this section was adapted from that of \cite{Neil:2022joj}.  The modified code that reproduces \cref{fig:fixed_N,fig:N_scaling_ga} is publically availble at \url{https://github.com/jwsitison/data_subset_variation}.}

Consider a simple toy problem where the ``true model" is a linear polynomial:
\begin{align}
f_{\T}(t)=1.80-0.53\left(1-\frac{t}{16}\right).
\end{align}
A set of $N$ samples are generated on $t\in\{1,2,\dots,15\}$ using $f_{\T}$ at each point multiplied by uncorrelated noise $1+\eta(t)$, where $\eta(t)$ is drawn from a Gaussian with mean $\bar{\eta}=0.0$ and variance $\sigma_{\eta}^2=1.0$. To be explicit, the mock data are drawn from $y(t)=(1+\eta(t))f_{\T}(t)$.

Our space of candidate data subsets will be labeled by the first data point $t_{\min}\in\{1,2,\dots,12\}$.  Each data subset will be fit with the following two polynomial models:
\begin{align}
f_0(t)&=a_0,\\
f_1(t)&=a_0+a_1\left(1-\frac{t}{16}\right).
\end{align}
We consider the case of moderately unconstrained parameter priors of a Gaussian with mean zero and width 10.

We use the model averaging procedures to determine the parameter estimate and error for $a_0$ in \cref{fig:fixed_N}, taking a data sample of size $N = 320$.  We use the following forms of the information criteria to determine the model weights:
\begin{align}
\label{eqn:baic_perf} \BAIC_{\mu,d_{\K}}^{\perf}=&-2\ln\pr(M_{\mu})+\hat{\chi}^2(\vec{a}^*)+2k-2d_{\K},\\
\label{eqn:bpic_subspace}
\BAIC_{\mu,d_{\K}}^{\subspace}=&-2\ln\pr(M_{\mu})+\hat{\chi}^2(\vec{a}^*)+2k-d_{\K}.
\end{align}
We will consider the case of flat priors over the model space, i.e., $\pr(M_0)=\pr(M_1)=1/2$, which makes the model prior term inconsequential.  We also report the $Q$-value of the fit (a Bayesian analogue of the $p$-value, see Appendix B of \cite{FermilabLattice:2016ipl}), which gives a measure of the fit quality.

Comparing the results of model averaging, we observe that the $\BAIC^{\perf}$ result has less uncertainty than $\BAIC^{\subspace}$. This is because $\BAIC^{\subspace}$ tends to favor fits with less data at larger $t_{\min}$, as is evident in the lower inset in \cref{fig:fixed_N}.  Since fits with lower statistics have higher uncertainty, weighting these fits in the $\BAIC^{\subspace}$ average inflates the error of the parameter estimate.  We note that this is a tendency, and not absolute; the $\BAIC^{\subspace}$ model weights do not satisfy the inequality \cref{eqn:thm-4.1}, which would require them to be strictly increasing with $t_{\rm min}$.  That inequality holds for the K-L divergence, whereas our numerical examples necessarily work with sample estimates of the BAIC, which only converge asymptotically to the K-L divergence.

We repeat the previous numerical test with several sample sizes $N\in\{40, 80, 160, 320, 640,$ $1280, 2400, 4800, 9600\}$; the final estimates for $a_0$ averaged over both data subsets ($t_{\min}\in\{1,\dots,12\}$) and models ($\mu\in\{0,1\}$) with $\BAIC^{\perf}$ and $\BAIC^{\subspace}$ are in \cref{fig:N_scaling_ga}.  As expected, the uncertainties in the $\BAIC^{\subspace}$ predictions are generally larger than those of $\BAIC^{\perf}$ for the same reasons discussed above.

From \cref{fig:N_scaling_ga}, we also see that the decreased variance of the $\BAIC^{\perf}$ compared to the $\BAIC^{\subspace}$ is not accompanied by increased asymptotic bias, which shows that this is not an instance of the bias-variance tradeoff \cite{Mehta2019}.  While the variance from the $\BAIC^{\subspace}$ is larger because it discards some data retained by the $\BAIC^{\perf}$, the retained data in either case is still drawn from the population and thus remains unbiased.

\section{Conclusion}
\label{sec:conclusion}

We have reviewed and compared two approaches of extending model variation criteria to data subset variation---the ``subspace" method introduced in \cite{Borsanyi:2020mff} and the ``perfect model" method introduced in \cite{Jay:2020jkz}.  While the arguments presented here are very general and apply to any K-L divergence--based information criteria (for example, those presented in \cite{Neil:2022joj}) and regression procedure, we have chosen to specialize to the AIC with least squares regression for simplicity.

While both formulas $\AIC^{\subspace}$ \cref{eqn:aic-subspace} and $\AIC^{\perf}$ \cref{eqn:aic-perfect} have similar qualitative features, namely a penalty for cutting away data, there are significant theoretical differences between the two methods that have important practical implications.  The subspace method involves discarding some of the data in a way that results in loss of information, and therefore tends to artificially inflate statistical uncertainty estimates.  The perfect model method avoids such a transformation by instead altering the fit models to include the cut data, which results in a stronger penalty for cutting away data.  We have shown this to be advantageous in the analytical and numerical examples considered.

For emphasis, the increased variance of $\AIC^{\subspace}$ compared to $\AIC^{\perf}$ is not accompanied by an increase in asymptotic bias (as in a typical bias-variance tradeoff).  The difference in statistical precision instead follows from the loss of information in $\AIC^{\subspace}$ associated with its tendency to use less of the available data compared to $\AIC^{\perf}$.  There is no sense in which either information criterion may be viewed as ``biased'' with respect to the other; they are simply estimating different quantities.

Although we advocate the use of the ``perfect model'' formula \cref{eqn:aic-perfect} (or equivalents for other information criteria) in all cases, we emphasize that the difference between this and the subspace formula is sub-leading to the effect of the fit $\chi^2$.  For example, results of model averaging will only differ significantly between the two formulas in cases where $\chi^2$ differences are small and the average is dominated by the data subset penalty term.  For any given statistical analysis, it could easily be the case that the difference in model-average results between the two formulas is negligibly small.

\begin{acknowledgments}

We thank Dan Hackett for useful discussions, and Laurent Lellouch for helpful conversations in the early stages of this work.  This work was supported in part by the U.~S.~Department of Energy (DOE), Office of Science, Office of High Energy Physics, under Award Number DE-SC0010005.

\end{acknowledgments}

\FloatBarrier
\bibliography{data-subset-variation}

\appendix

\section{Asymptotic results for the K-L divergence} \label{sec:app-asymptotic-kl}

Here, we consider the evaluation of the first term in the K-L divergence \cref{eqn:model-variation-KL} for a given data set of dimension $d$:
\beq
E_z[\ln \pr_{M_{\T}}(z)] = \int dz\ \pr(z) \ln \pr(z).
\eeq
To proceed, we need more information about the true distribution.  If we assume that our data are generated from a Gaussian random process, then the true distribution will be multivariate Gaussian:
\beq
\pr(z) \rightarrow \frac{1}{(2\pi)^{d/2} (\det \Sigma)^{1/2}} \exp \left( -\frac{1}{2} (z-\mu_{\T})^T \Sigma_{\T}^{-1} (z-\mu_{\T}) \right),
\eeq
where $\mu_{\T}$ and $\Sigma_{\T}$ are the true mean and true covariance, respectively.  (More generally, we could work with a sample estimator of $\pr_{M_{\T}}(z)$ based on the sample means $\bar{y}$, so that the central limit theorem applies; this would lead to the same multivariate Gaussian form.)

Using this form of the distribution, and introducing the shorthand notation
\beq
(z-\mu_{\T})^T \Sigma_{\T}^{-1} (z-\mu_{\T}) \equiv \chi_{\T}^2(z)
\eeq
we have
\begin{align}
\nonumber E_z[\ln \pr_{M_{\T}}(z)] = \frac{1}{(2\pi)^{d/2} (\det \Sigma_{\T})^{1/2}} \int dz\ &\left[ -\frac{d}{2} \ln (2\pi) - \frac{1}{2} \ln \det \Sigma_{\T} - \frac{1}{2} \chi_{\T}^2(z) \right]\\
&\cdot \exp\left(-\frac{1}{2}\chi_{\T}^2(z)\right).
\end{align}
The first two terms are simply expectation values of constants, which are trivial to evaluate.  For the third term, we can use a standard differentiation trick to simplify:
\begin{align}
-\frac{1}{2} \int dz\ \chi_{\T}^2(z) \exp(-\chi_{\T}^2(z)/2) &= \left. \int dz\ \frac{d}{d\alpha} \exp(-(\alpha/2) \chi_{\T}^2(z))\right|_{\alpha=1} \\
&= \frac{d}{d\alpha} \left[ (2\pi/\alpha)^{d/2} (\det \Sigma_{\T})^{1/2} \right]_{\alpha = 1} \\
&= -\frac{d}{2} (2\pi)^{d/2} (\det \Sigma_{\T})^{1/2}.
\end{align}
Plugging back in, we thus have
\beq
E_z[\ln \pr_{M_{\T}}(z)] = -\frac{d}{2} \ln (2\pi) - \frac{1}{2} \ln \det \Sigma_{\T} - \frac{d}{2}.
\eeq
The first two terms are essentially just the normalization of the probability distribution $\pr(z)$.  If we return to the complete K-L divergence \cref{eqn:model-variation-KL}, the second term is
\beq
-E_z [\ln \pr(z | M_\mu, \vec{a}^*) ] = -\int dz\ \pr(z) \ln \left[\frac{1}{(2\pi)^{d/2} (\det \Sigma)^{1/2}} \exp \left( -\frac{1}{2} \chi^2(z | M_\mu, \vec{a}^*) \right) \right].
\eeq
We can't proceed further with the term that contains the plug-in likelihood $\chi^2(z | M_\mu, \vec{a}^*)$, but we see that once again this second term will contain the analogues of the first two terms in $E_z[\ln \pr_{M_{\T}}(z)]$,
\beq
-E_z [\ln \pr(z | M_\mu, \vec{a}^*) ] = +\frac{d}{2} \ln(2\pi) + \frac{1}{2} \ln \det \Sigma + ...
\eeq
The data covariance matrix $\Sigma$ is an unbiased estimator for the true covariance matrix $\Sigma_{\T}$, so the difference $(\ln \det \Sigma_{\T} - \ln \det \Sigma)$ will vanish in the asymptotic limit (number of samples $N \rightarrow \infty$).  Thus, including the conventional factor of $+2N$ for conversion to the AIC, see \cref{eqn:aic-definition} (with an extra minus sign since we are now keeping a contribution from the first term in the K-L divergence \cref{eqn:model-variation-KL}), the extra contribution that does not cancel is precisely $-Nd$, which becomes $-Nd_{\K}$ for the subset method of \cref{subsec:aic-subspace} as we only evaluate this term on the kept data.

\section{AIC and degrees of freedom \label{sec:AIC-dof}}

In this section, we consider rewriting the two different AIC formulas in terms of the ``number of degrees of freedom'' $\Ndof \equiv d_{\K} - k$.  This is precisely the namesake quantity for the $\chi^2$ statistic for fits to the kept data, from a frequentist (or Bayesian with very diffuse priors) point of view.  In terms of this quantity, dropping any constant terms proportional to $d$, we have
\begin{align}
\AIC_{\mu, d_{\K}}^{\subspace} &= \Ndof \left( \hat{\chi}_{\K}^2(\vec{a}^*)/\Ndof-1 \right) + k, \\
\AIC_{\mu, d_{\K}}^{\perf} &= \Ndof \left(\hat{\chi}_{\K}^2(\vec{a}^*)/\Ndof-2\right).
\end{align}
At first glance, this reformulation appears to show problematic behavior for the $\AIC^{\perf}$ formula.  In the limit that $\Ndof$ is very large, the value of $\hat{\chi}_{\K}^2(\vec{a}^*)/\Ndof$ is expected to be close to 1 for a fit which describes the data well.  Fits with $\hat{\chi}_{\K}^2(\vec{a}^*)/\Ndof > 1$ are very bad (extremely low $p$-value) if $\Ndof$ is sufficiently large.  For such bad fits, the first formula gives $\AIC^{\subspace} > 0$, but the second formula gives $\AIC^{\perf} < 0$.  This, apparently, allows obviously bad fits with large $\Ndof$ to be favored in model variation relative to marginally good fits with small $\Ndof$, for which $\hat{\chi}_{\K}^2(\vec{a}^*)/\Ndof \sim 2$ could have a larger $p$-value.

To be more concrete, let us consider a comparison of two hypothetical fits using both formulas.  Fit A has a small number of degrees of freedom $\Ndof^A \equiv n$, and has what would normally be considered a good fit with $\hat{\chi}_{\K,A}^2 / \Ndof^A \approx 1$.  On the other hand, fit B has a large number of degrees of freedom $\Ndof^B \equiv N \gg n$, and is a poor fit, $\hat{\chi}_{\K,B}^2 / \Ndof^B \approx 2$.  The resulting information criterion values are:
\begin{align}
\AIC_{A,d_{\K}}^{\subspace} &\approx 0 + k, \quad &\AIC_{B,d_{\K}}^{\subspace} &\approx N + k, \\
\AIC_{A,d_{\K}}^{\perf} &\approx -n, \quad &\AIC_{B,d_{\K}}^{\perf} &\approx 0.
\end{align}
In both cases, the ``good'' fit A is favored over ``bad'' fit B.  However, the relative model probability is $p_{B}^{\perf}/p_{A}^{\perf} \sim e^{-n}$ in the perfect model case, disfavoring model B slightly, while in the subset construction $p_{B}^{\subspace}/p_{A}^{\subspace} \sim e^{-N}$ and the bad fit is disfavored very strongly.

Does this result indicate some inconsistent behavior for the perfect-model approach?  Although it appears troubling at first glance, we argue that the behavior of $\AIC^{\perf}$ in this example is entirely reasonable.  In the context of data subspace selection, the large difference in $\Ndof$ between fit A and fit B must come from a large difference in the amount of data cut away.  This is why ``good'' fit A is only slightly disfavored by the perfect model construction---it secretly has a large number of additional ``fit parameters'' that are used on the cut portion of the data.  (From the converse point of view, $\AIC^{\subspace}$ is overweighting fit A by not penalizing heavily enough for the large amount of data that has been cut away.)  We can emphasize this interpretation by rewriting the perfect model AIC as follows:
\begin{align}
\AIC_{\mu, d_{\K}}^{\perf} &= \Ndof \left(\hat{\chi}_{\K}^2(\vec{a}^*)/\Ndof-1\right) + k - d_{\K} = \AIC_{\mu, d_{\K}}^{\subspace} - d_{\K}
\end{align}
In this form, it is clear that the difference is simply the magnitude of the data-cutting penalty.  Both forms of AIC will incur an equal goodness-of-fit penalty for fit B for having $\chi^2 / \Ndof \approx 2$, but the perfect model increases the overall weight of fit B due to the significantly larger $d_{\K}$ compared with fit A.

\end{document}